\documentstyle[prl,aps,epsf,twocolumn]{revtex}

\newcommand{\be}{\begin{equation}}
\newcommand{\ee}{\end{equation}}
\newcommand{\bea}{\begin{eqnarray}}
\newcommand{\eea}{\end{eqnarray}}
\newcommand{\Sv}{{\bf S}}
\newcommand{\xv}{{\bf x}}
\newcommand{\yv}{{\bf y}}
\newcommand{\Hvs}{{\bf H}_s}
\newcommand{\nv}{{\bf n}}
\newcommand{\mvs}{{\bf m}_s}
\newcommand{\nvt}{\tilde{{\bf n}}}
\newcommand{\Jv}{{\bf J}}
\newcommand{\pa}{\partial}
\newcommand{\non}{\nonumber}

\begin{document}
\draft
\twocolumn[\hsize\textwidth\columnwidth\hsize\csname  
@twocolumnfalse\endcsname
\author{E. Ercolessi$^{(1)}$, G. Morandi$^{(1)}$, P. Pieri$^{(2)}$, and 
M. Roncaglia$^{(1)}$}
\address{(1) Dipartimento di Fisica, Universit\`a di Bologna, INFN and INFM, 
V.le Berti Pichat 6/2, I-40127, Bologna, Italy\\
(2) Dipartimento di Matematica e Fisica and INFM, 
Universit\`{a} di Camerino, I-62032 Camerino, Italy}
\title{Integer-spin Heisenberg Chains in a Staggered Magnetic Field.\\
A Nonlinear $\sigma$-Model Approach.}

\maketitle
\hspace*{-0.25ex}
\begin{abstract}
We present here a nonlinear sigma-model (NL$\sigma$M) study of a spin-1
antiferromagnetic Heisenberg chain in an external commensurate staggered
magnetic field. We find, already at the mean-field level, excellent agreement
with recent and very accurate Density Matrix Renormalization Group (DMRG)
studies, and that up to the highest values of the field for which a comparison
is possible, for the staggered magnetization and the transverse spin gap.
Qualitative but not quantitative agreement is found between the NL$\sigma$M
predictions for the longitudinal spin gap and the DMRG results.   The origin
of the discrepancies is traced and discussed. Our results allow for extensions
to higher-spin chains that have not yet been  studied numerically, and the
predictions for a spin-2 chain are presented and discussed. Comparison is
also made with previous theoretical approaches that led instead to predictions
in disagreement with the DMRG results.
\end{abstract}
\vskip2pc] \narrowtext

One and quasi-one-dimensional magnets (spin chains and ladders) have become
the object of intense analytical, numerical and experimental studies since
Haldane put forward \cite{1} his by now famous ``conjecture'' according to which
half-odd-integer spin chains should be critical with algebraically decaying
correlations, while integer-spin chains should be gapped with a disordered
ground state \cite{2,3}. Haldane's conjecture has received strong experimental
support from neutron scattering experiments on quasi-one-dimensional spin-1
materials such as NENP and Y$_2$BaNiO$_{5}$ \cite{4}. 
It is by now well established that pure one-dimensional Haldane systems (i.e.
integer-spin chains) have a disordered ground state with a gap to a degenerate
triplet (for spin-1 systems) of magnon excitations. There is also general
consensus on the value of: $\Delta_{0}=0.41048(2)J$ (with $J$ the exchange
constant) for the gap in spin-1 chains, that has been obtained by both
Density Matrix Renormalization Group (DMRG) \cite{7} and finite-size exact
diagonalization \cite{8} methods.

The effects on Haldane systems of external magnetic fields have also been the
object of intense studies, again both experimental, numerical and analytical.
A uniform field induces a Zeeman splitting of the degenerate magnon triplet,
with the Haldane (gapped) phase remaining stable, with zero magnetization, up
to a lower critical field $H_{c_{1}}=\Delta_{0}$, when one of the magnon
branches becomes degenerate with the ground state and a Bose condensation of
magnons takes place \cite{9,10}. At higher fields the system enters into a
gapless phase and magnetizes, with the magnetization reaching saturation at an
upper critical field $H_{c_{2}}=4SJ$. 

In view of the underlying, short-range antiferromagnetic (AFM) ordering that
is present in the Haldane phase, the study of the effects of a
{\it staggered} magnetic field appears to be even more interesting. Of
course static, staggered fields cannot be manufactured from the outside.
However, recently a class of quasi-one-dimensional compounds has been
investigated that can be described as spin-1 chains acted upon by an
effective internal staggered field. Such materials have Y$_{2}$BaNiO$_{5}$ as
the reference compound and have the general formula R$_{2}$BaNiO$_{5}$ where R
($\neq$ Y) is one of the magnetic rare-earth ions. The reference compound is
found to be highly one-dimensional, hence a good realization of a Haldane-gap
system (remember that the Ni$^{2+}$ ions have spin $S=1$) \cite{15}. 
The magnetic R$^{3+}$ ions are positioned between neighboring Ni chains and weakly coupled with them. Moreover, they order antiferromagnetically below a certain N\'{e}el temperatute $T_{N}$ (typically: 
16 K$\lesssim T_{N}\lesssim80$ K \cite{15}). 
This has the effect of imposing an effective staggered (and
commensurate) field on the Ni chains. The intensity of the field can be
indirectly controlled by varying the temperature below $T_{N}$. \ The
staggered field lifts partially the degeneracy of the magnon triplet, leading
to different spin gaps in the longitudinal (i.e. parallel to the field) and
transverse channels. Neutron scattering experiments on samples of
Nd$_{2}$BaNiO$_{5}$ and Pr$_{2}$BaNiO$_{5}$ \cite{16} show an increase of the
Haldane gap as a function of the staggered field. Experimental results have
also been obtained for the staggered magnetization \cite{17}.

Recently, an extensive DMRG study of an $S=1$ Heisenberg chain in a
commensurate staggered field has appeared \cite{18}. The authors in Ref. \cite{18}
have obtained accurate results for the staggered magnetization curve,
the longitudinal and transverse gaps and the static correlation functions
(both longitudinal and transverse). They found however a strong disagreement
in the high-field regime between their results and previous theoretical
approaches \cite{19,20} whose starting point was a mapping of the chain onto a
nonlinear sigma-model, and this has led them to openly question the entire
validity of the NL$\sigma$M approach, at least in the high-field regime.

This strong criticism has partly motivated our study of the same model as in
Ref. \cite{18}. We will actually show that an accurate treatment of the 
NL$\sigma$M does indeed lead to an excellent agreement between the analytical 
and the
DMRG results both for the magnetization and the transverse gap for all
values of the staggered field. Some discrepancies are instead present for the
longitudinal gap between our results and the data of Ref. \cite{18}, but our
identification of the latter relies on an approximation that, as we shall
argue in the concluding part of this Letter, may be questionable in the
presence of an external (staggered) field. After presenting the derivation of
our results, we will also discuss what are the reasons for the disagreement
between previous theoretical approaches and the DMRG results.

We start from the following Hamiltonian for a Heisenberg chain coupled
to a staggered magnetic field:
\be
H=\sum_{i=1}^N \left[ J \Sv_i \cdot \Sv_{i+1} + 
\Hvs \cdot (-1)^i \Sv_i \right]
\label{ham}
\ee 
where $\Sv_i^2=S(S+1)$ (we set $\hbar=1$ from now on), the staggered
magnetic field $\Hvs$ includes Bohr magneton and gyromagnetic factor, and
$J>0$ is the exchange constant. In this Letter we will consider the 
case of ${\em integer}$ spin $S$. Under this assumption, the Haldane mapping
of the Heisenberg chain onto a (1+1) nonlinear $\sigma$-model  
can be readily obtained\cite{20}. The topological term is absent since the
spin $S$ is integer, while the staggered magnetic field couples linearly to 
the NL$\sigma$M field $\nv$. 
The euclidean NL$\sigma$M Lagrangian is thus given by\cite{20}:
\be
{\cal L}_E=\frac{1}{2 g}\left[ \frac{1}{c}\left(\frac{\pa \nv}{\pa \tau}
\right)^2+ c\left(\frac{\pa \nv}{\pa x}\right)^2- 2 g S \, \nv\cdot \Hvs 
\right]\;\;,
\label{leucl}
\ee
where the NL$\sigma$M field $\nv$ is a three-component unit vector pointing in the direction of the local staggered magnetization. The bare 
coupling constant $g$ and spin-wave velocity $c$ are related to the 
parameters of the Heisenberg Hamiltonian by  $g=2/S$ and $c=2 J S a$, 
with $a$ the lattice spacing.
The constraint $\nv^2=1$ can be taken into account in the path-integral expression for the partition function by writing 
${\cal Z}= \int\left[{\cal D}\nv\right]
\left[\frac{{\cal D}\lambda}{2\pi}\right] e^{-S_E}$,  
where $S_E=\int_0^{L}d x\int_0^\beta d \tau {\cal L}(x,\tau)$ 
($L=N a$ being the total length of the chain) and
\be
{\cal L}(x,\tau)={\cal L}_E(x,\tau)- i \lambda(x,\tau)[\nv(x,\tau)^2-1]
\;\;.
\ee
The functional integration over the field $\lambda(x,\tau)$ implements
the constraint $\nv^2=1$.

The euclidean action $S_E$ is quadratic in the field $\nv$ which can be 
integrated out exactly. To this end, we introduce the notation 
$\xv=(x,\tau)$ and write
\bea
S_E &=&\int d \xv \, d \xv' \, K(\xv,\xv')\nv(\xv)\cdot\nv(\xv')\non\\
&-&S\Hvs\cdot\int d\xv \, \nv(\xv) + i\int d\xv \,\lambda(\xv)\;\;,
\label{sq}
\eea
where the kernel $K$ is given by: 
\be
K(\xv,\xv')=-\frac{1}{2 g c}[ c^2\pa^2_x+\pa^2_{\tau}+ 2 i g c \lambda(\xv)]
\delta(\xv-\xv')\;\;. \label{ker}
\ee
The square in equation (\ref{sq}) can be completed by performing the linear 
shift $\nv(\xv)\to \nv'(\xv) = \nv(\xv)-\bar{\nv}(\xv)$, where
\be
\bar{\nv}(\xv)=\frac{S}{2}\Hvs\int d\xv'K^{-1}(\xv,\xv')\;\;\;,
\ee
and $K^{-1}$ is the inverse of the kernel (\ref{ker}).  
After the integration over the shifted field we obtain the effective action
for the field $\lambda$
\bea
S(\lambda)&=&\frac{3}{2}{\rm  Tr} \ln K + i \int d \xv \lambda(\xv)
\non\\
&-& \frac{1}{4} S^2 \Hvs^2\int d \xv \, d \xv' \, K^{-1}(\xv,\xv')\;\;\; 
\eea
The partition function is now given by ${\cal Z} = \int [\frac{{\cal D}\lambda}{2\pi}]
e^{-S(\lambda)}$ and can be calculated by a saddle-point approximation
for the integral over $\lambda$.
We look for a static and constant saddle-point solution: the kernel $K$
depends in this case only on the relative coordinates 
$K(\xv,\xv')=K(\xv-\xv')$.
The equation $\frac{\pa S}{\pa\lambda}=0$ reads then:
\be
\frac{3}{2} K^{-1}(\xv=0)=1-\frac{S^2 \Hvs^2}{4} \left[\int d \xv K^{-1}
(\xv) \right]^2\;\;\;. \label{sadd}
\ee
It is convenient to work now in Fourier space, which is defined according to 
\be
\nvt(q,n)=\int_0^L dx \int_0^{\beta} d \tau e^{-i(q x - \Omega_n\tau)}\nv(x,\tau)
\ee
(the frequencies $\Omega_n=2\pi n/\beta$ are Matsubara Bose frequencies), and 
similarly for the field $\lambda$ and kernel $K$. 
The saddle-point equation becomes:
\bea
\frac{3}{2 (\beta L)}\sum_{q,n}\tilde{K}^{-1}(q,n)&=&1-\frac{S^2 \Hvs^2}{4} 
\left[\tilde{K}^{-1}(q=0,n=0)\right]^2
\label{sp1}\\
\tilde{K}^{-1}(q,n)&=&\frac{ 2 g c }{\Omega_n^2 + c^2 q^2 - i \lambda 2 g c }
\label{Kt}
\;\;\;.
\eea
We now make the assumption $- i \lambda 2 g c = c^2/\xi^2$,
with $\xi^2$ real and positive. Under this assumption, that  will be 
consistently verified shortly below, the sum over frequencies in 
Eq.~(\ref{sp1}) can be performed by standard techniques. 
 
By taking the thermodynamic and zero-temperature limit ($L,\beta\to\infty$) 
we finally obtain the following equation for $\xi$:
\be
\frac{3 g}{2 \pi}\ln\left[\Lambda \xi+ \sqrt{1+(\Lambda\xi)^2}\right]=
1-\frac{S^2 g^2 \Hvs^2}{c^2}\xi^4\;\;\;,
\label{eq1}
\ee
where we have introduced the ultraviolet cutoff $\Lambda$ to 
regularize the momentum integration. The staggered magnetization 
(in units of $S$) will be determined instead by the equation for the linear 
shift, which now reads
\be
\mvs=\langle\bar{\nv}\rangle=\frac{S\Hvs}{2}\tilde{K}^{-1}
(q=0,n=0)=\frac{g S \xi^2}{c}\Hvs\;\;\;.
\label{eq2}
\ee
In order to get rid of the ultraviolet cutoff $\Lambda$ we observe that 
from Eq.~(\ref{Kt}) it follows that, when $\Hvs=0$, the Green functions
for the field $\nv$ are given by
\bea
G_{\alpha,\beta}(q,n)&=&\langle \tilde{n}_{\alpha}(q,n)\tilde{n}_{\beta}(-q,-n)
\rangle 
=\delta_{\alpha,\beta}\frac{1}{2}\tilde{K}^{-1}(q,n)\non\\&=&
\delta_{\alpha,\beta}\frac{g c}{\Omega_n^2 + c^2 q^2 + c^2/\xi^2}
\;\;\;,
\label{green}
\eea  
with $\xi=\sinh(2\pi/3 g)/\Lambda$, as obtained from Eq. (\ref{eq1}) for 
$\Hvs=0$. From the structure of the Green functions (\ref{green})
it follows that the excitations are gapped, with the value 
\be
\Delta_0=\Delta(\Hvs=0)=\frac{\Lambda c}{\sinh(2\pi/3 g)} \label{eq3}
\ee  
for the zero-field Haldane gap\cite{1} $\Delta_0$. 
The Haldane gap is the only free parameter in our theory, 
that we will assume from previous numerical estimates \cite{7,8}. 
In this way we fix the cutoff.
Eqs.~(\ref{eq1}) and (\ref{eq2}) (with the addition of Eq.~(\ref{eq3})) 
provide our staggered-magnetization curve. Eq.~(\ref{eq1}) can be solved 
analytically for large or small values $\Hvs$, while in general it has to
be solved numerically (albeit with no effort).

For small $\Hvs$ we obtain
\be
\mvs=\frac{g S c}{\Delta_0^2}\Hvs\left(1 -\frac{4 \pi/3}{\tanh(2\pi/3 g)}
\frac{c^2g S^2}{\Delta^4}H_s^2 + {\rm O}(H_s^4)\right)\;\;\;.
\ee
The zero-field staggered susceptibility is then given by
\be
\chi_s(0)=\left. \frac{d m_s}{d H_s}\right|_{H_s=0}
=\frac{g S c}{\Delta_0^2}= \frac{4 J S}{\Delta_0^2}
\ee
For $S=1$, by taking $\Delta_0=0.41048 J$ from Ref.\cite{7}, we obtain
$\chi_s(0)=23.74/J$ which agrees fairly well with Monte-Carlo result 
$\chi_s(0)=21(1)/J$\cite{sak} and DMRG calculation $\chi_s(0)=18.5/J$
\cite{18}. For $S=2$, we take $\Delta_0=0.0876 J$ from Ref.\cite{18a}  
and get $\chi_s(0)=1043/J$, which should be checked by new numerical 
investigations. 

For large $\Hvs$ the staggered magnetization is instead given by
\be
m_s= 1 - \frac{A}{\sqrt{H_s}} + \frac{1}{2}\frac{A^2}{H_s} +
{\rm O}(A^3/H_s^{3/2})
\label{largh}
\ee
\begin{figure}[t]
\begin{center}
\epsfxsize=6cm \epsffile{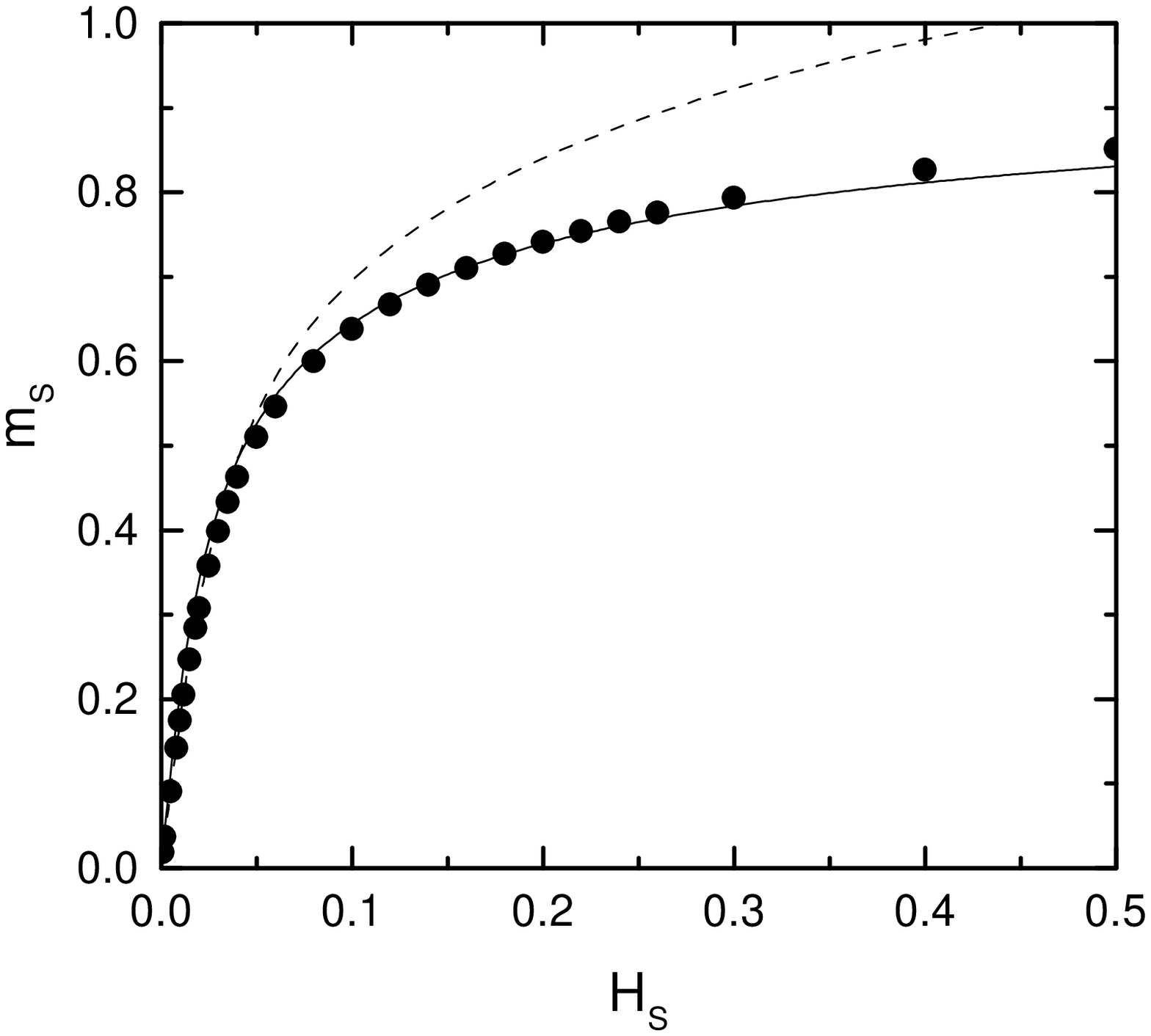}
\narrowtext
\caption[]{The staggered-magnetization curve for $S=1$. 
Our results (solid line) are compared with DMRG data (circles) and with the NL$\sigma$M results (dotted line) of Refs.\cite{18} and \cite{20}, 
respectively.}
\end{center}
\label{fig1}
\end{figure}
\noindent
where $A=\Delta_0\frac{3}{4\pi}\frac{\sinh(\pi S/3)}{S(J S)^{1/2}}$.
It is obvious from Eq. (\ref{largh}) that the magnetization saturates to its maximum
value of $m_{s}=1$ only asymptotically for $H_{s}\to\infty$. 
This is consistent with the fact that it is only in that asymptotic limit that a fully
polarized N\'{e}el state becomes an exact eigenstate of the original
Heisenberg Hamiltonian (\ref{ham}), as well as with the fact that our mean-field
approach keeps track of the NL$\sigma$M constraint, albeit only on the
average.

In Figure 1 we compare, for $S=1$, our staggered-magnetization curve with  
DMRG data~\cite{18} and with the NL$\sigma$M treatment of Ref.~\cite{20}.
The magnetization curve was obtained by solving numerically Eq.~(\ref{eq1})
(with  $\Lambda$ given by Eq.~(\ref{eq3})) and by inserting the resulting 
$\xi$ into Eq.~(\ref{eq2}). 
As already anticipated, the agreement with the DMRG data is excellent over
the whole range of fields.

We have also solved Eq.~(\ref{eq1}) for $S=2$, by fixing again $\Lambda$ 
via Eq.~(\ref{eq3}) with the value, specific to $S=2$, $\Delta_0=0.0876 
J$\cite{18a}. At present, the staggered-magnetization curve for $S=2$ 
has not yet been obtained by numerical methods. Given the excellent agreement
between our curve and numerical data for $S=1$, we expect that our 
magnetization curve (shown in Figure 2) will agree with future numerical
investigations also in the case $S=2$. 
Indeed, on general grounds, Haldane's mapping is expected to work even better 
when the value of the spin $S$ is increased. 
Note that the magnetization increases faster with $H_s$ 
(or, equivalently, the staggered susceptibility becomes larger) when the spin 
$S$ is increased. This is consistent with the fact that, when $S\to\infty$, 
the system approaches the classical one, which is critical at $T=0$ and
$H_s=0$. 
Note also that the value of the zero-field Haldane gap roughly
sets the scale of the staggered-magnetization curve.
\begin{figure}[t]
\hspace{1.cm}
\epsfxsize=6cm \epsffile{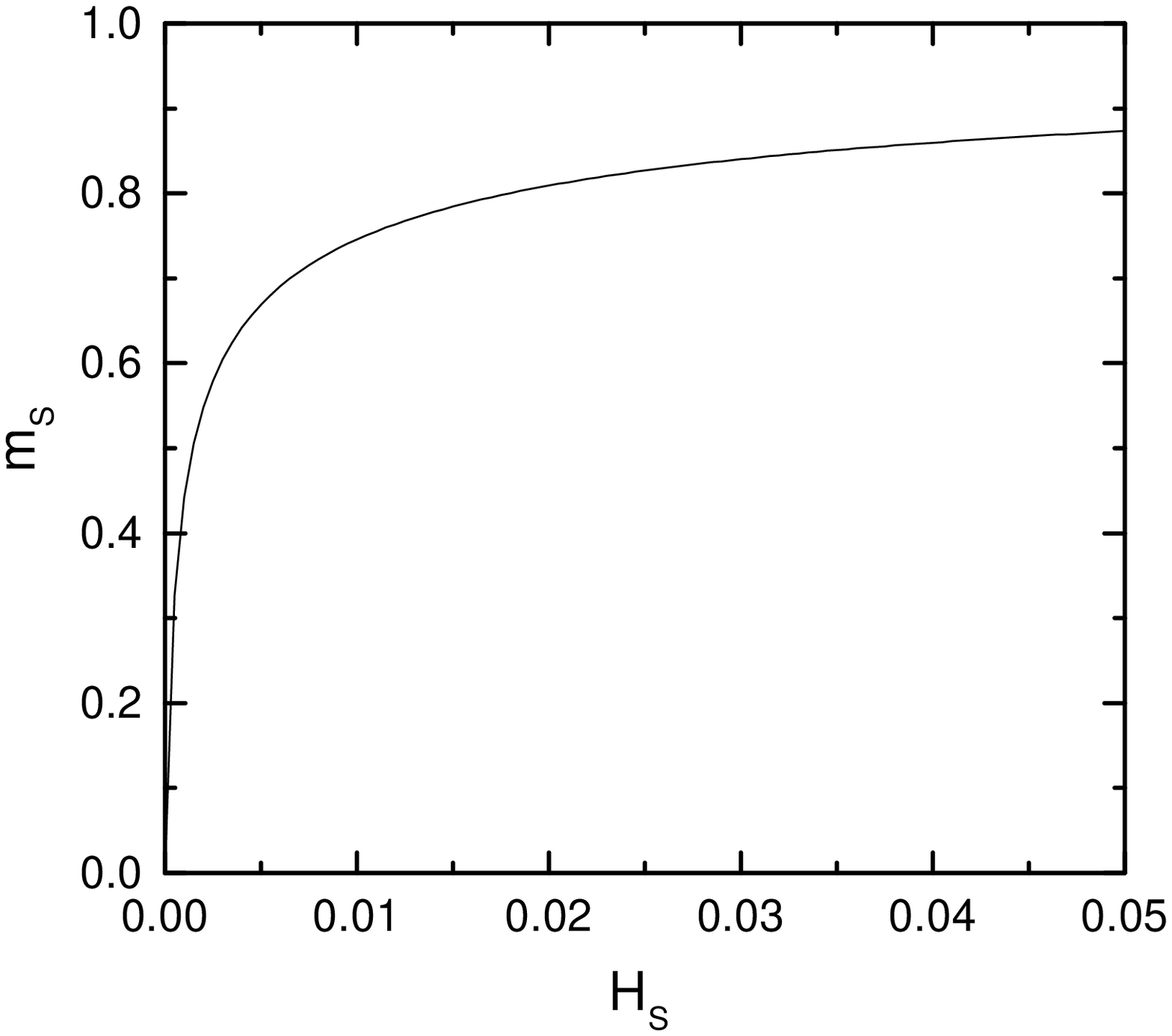}
\narrowtext
\caption[]{The staggered-magnetization curve for S=2.}
\label{fig2}
\end{figure}

Introducing an external space and time-dependent source field 
$\mathbf{J=J(x)}$ and promoting the partition function ${\cal Z}$ to a generating
functional ${\cal Z}[\Jv]$ obtained by replacing in the original
path-integral the Euclidean Lagrangian of Eq.~(\ref{leucl}) with:
\be
{\cal L}[\Jv]=\frac{1}{2 g}\left[ \frac{1}{c}\left(\frac{\pa \nv}{\pa \tau}
\right)^2+ c\left(\frac{\pa \nv}{\pa x}\right)^2\right]- \nv(\xv)\cdot 
\Jv(\xv) 
\;\;,
\label{lj}
\ee
the connected Green functions for the staggered field $\nv$:
\be
G_c^{\alpha,\beta}(\xv,\xv')=\langle {\rm T_{\tau}}[n^{\alpha}(\xv) 
n^{\beta}(\xv')]\rangle - \langle n^{\alpha}(\xv)\rangle \langle 
n^{\beta}(\xv')\rangle
\ee
can be obtained by double functional differentiation with respect to the
source field as:
\be
G_c^{\alpha,\beta}(\xv,\xv')=
\left. \frac{\delta^2 \ln {\cal Z} [\Jv]}{\delta \Jv^{\alpha}(\xv) \delta 
\Jv^{\beta}(\xv')}\right|_{\Jv= S \Hvs}\;\;\;.
\label{func}
\ee
The saddle-point analysis can be performed in this case as well. The
saddle-point equation reads now:
\bea
\frac{3}{2} K^{-1}(\xv,\xv)&=&1-\frac{1}{4} \int  
d \yv d \yv'  K^{-1}(\yv,\xv) K^{-1}(\xv,\yv')\non\\
&\times & \Jv(\yv) \cdot \Jv(\yv')\;\;\;.
\label{sadp}
\eea
This equation will yield in general a space and time-dependent saddle point
$\lambda_{sp}=\lambda_{sp}(\xv)$, which will reduce however to a
constant for $\Jv=const.=S \Hvs$ when, as it can be checked
easily, Eq. (\ref{sadp}) reproduces our previous Eq. (\ref{sadd}). 
What matters here is the
fact that the saddle point will depend {\it quadratically} on the source field. 
From this it follows that the {\em transverse} Green functions 
$G_{xx}(\xv,\xv')=G_{yy}(\xv,\xv')$ do not get any additional contribution from the 
dependence of $\lambda_{{\rm sp}}$ on $\Jv$ (we assume here $\Hvs= {\hat z} 
H_s$) and can be calculated by taking $\Jv=S \Hvs$ from the beginning.
The transverse Green functions are thus still given by the same 
expression valid for $H_s=0$, Eq.~(\ref{green}) where now $\xi$ depends
on $\Hvs$ via Eq. ({\ref{eq1}). The {\em transverse} gap will be accordingly
$\Delta_{\perp}(\Hvs)= \frac{c}{\xi(\Hvs)}$, and in Figure 3 we compared the transverse gap as calculated by this relation with the one 
obtained by DMRG data. The agreement is again excellent. 
  
\begin{figure}[t]
\hspace{1.cm}
\epsfxsize=6cm \epsffile{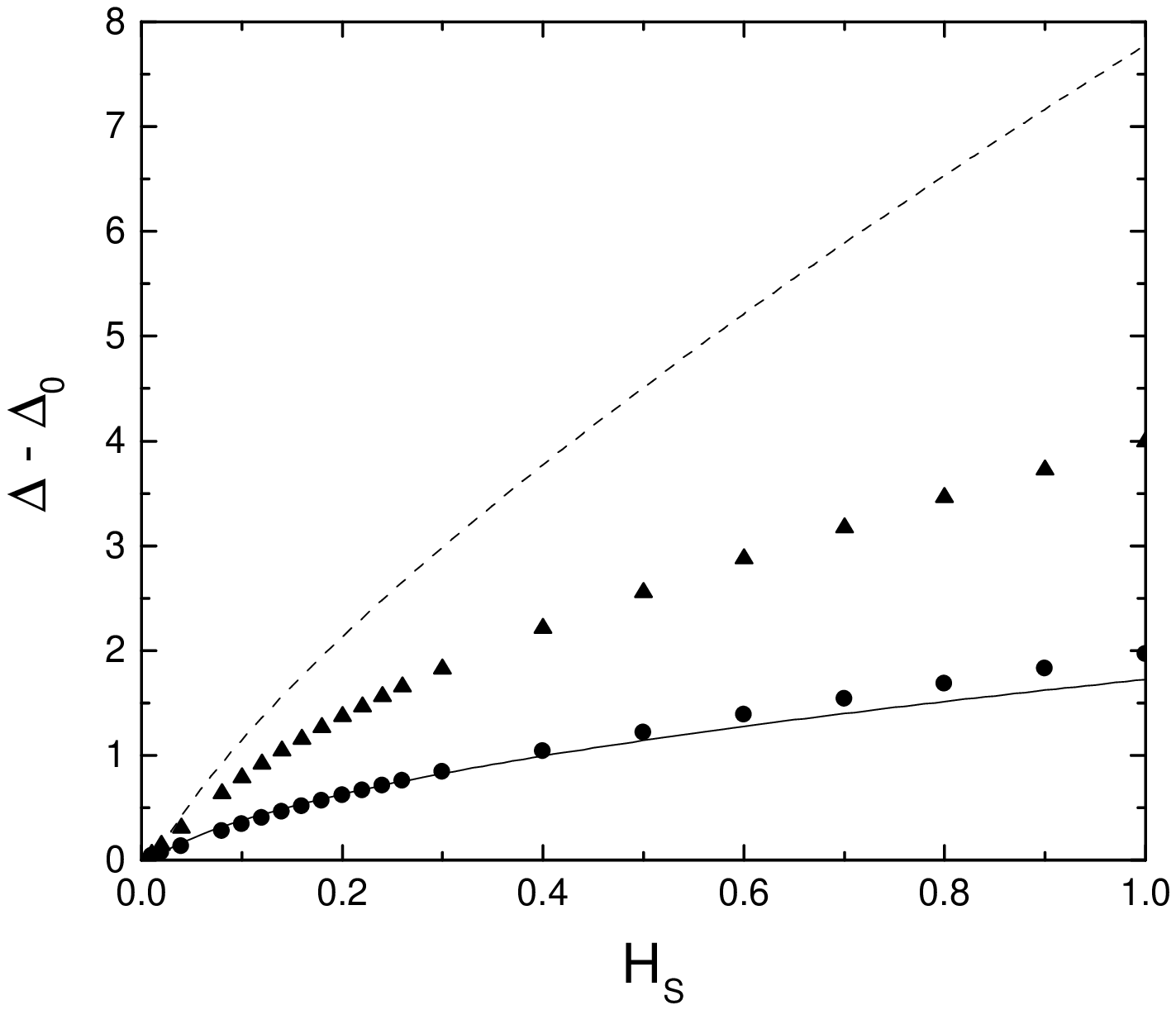}
\narrowtext
\caption[]{Plots of the transverse (solid curve) and longitudinal 
(dashed curve) gap predicted by the NL$\sigma$M approach, compared with 
the corresponding DMRG data (circles and triangles, respectively).}
\label{fig3}
\end{figure}

In the {\em parallel} channel, on the other hand, the dependence of $\lambda_{{\rm sp}}$ 
on $\Jv$ yields additional contributions to the Green functions, which do not 
allow us to calculate the longitudinal Green functions explicitly. 
To obtain an estimate for the longitudinal gap we resorted to the so called
``single-mode approximation''\cite{20}, that is we assumed the gap 
and susceptibilities to be connected by the same law both in the transverse and 
parallel channel. It can be readily shown that for an isotropic system, for 
which the magnetization is directed along the external field,  
$\chi_{\perp}=\frac{\pa m_s^x}{\pa H_s^x}= m_s/H_s$, while 
$\chi_{\parallel}=\frac{\pa m^z_s}{\pa H^z_s}=\frac{d m_s}{d H_s}$. From 
Eq.~(\ref{eq2}) we thus
find $\chi_{\perp}= S g \xi^2/c=S g c/(\Delta_{\perp})^2$.
If we assume the same relation to hold
also in the parallel channel, $\Delta_{\parallel}$ can be calculated 
as a derived quantity from the magnetization curve by using 
$\Delta_{\parallel}=(S g c /\chi_{\parallel})^{1/2}$.
Our data for $\Delta_{\parallel}$ are also shown in Figure 3. 
The agreement with the DMRG data is definitely worse in this case, and
this may attributed to a shortcoming of the single-mode approximation in the
longitudinal channel. 
The complete structure of the Green functions along the lines outlined
above in Eqs. (\ref{lj}) and (\ref{func}) is being presently undertaken, and more extended and
complete results will be published elsewhere \cite{23}.

In conclusion, we come to a brief comparison between our approach and
that of Refs. \cite{19} and \cite{20}. 

What the two approaches have in common is the starting point, 
namely the mapping of the original
problem onto a NL$\sigma$M, slightly modified by the external
staggered field. In the abovemntioned references, however, the authors
resolved to relaxing the NL$\sigma$M constraint by replacing it with a
polynomial self-interaction of the $\nv$-field, keeping terms in the
expansion of the interaction up to eight order. This leads to a (generalized)
Ginzburg-Landau-type theory with up to six free parameters which the authors
took from previous numerical and Renormalization-Group studies.
Their results for the staggered magnetization saturate
at a finite value of the staggered field (see Figure 1), 
which indicates that softening the NL$\sigma$M constraint is inappropriate 
at high fields. 
Also, their results for the gaps deviate
badly from the DMRG results. 
All this has been evidenced in Ref. \cite{18} 
and justifies the negative comments of the authors in that reference, that however
should be addressed more to the additional approximations adopted in Refs. 
\cite{19} and \cite{20} than to the NL$\sigma$M approach itself. 

\acknowledgements
The authors are grateful to Prof.Yu Lu and Dr.Jizhong Lou for letting them
have access to the files of their DMRG data.

\end{document}